# The magnetic behaviour of $Dy_2Ir_2O_7$ – beyond the mean field approximation

K. Vlášková[1], M. Diviš[1], M. Klicpera[1]

[1]*Charles University, Faculty of Mathematics and Physics, Department of Condensed Matter Physics, Ke Karlovu 5, 121 16 Prague 2, Czech Republic*

**Abstract**

The magnetic properties of a pyrochlore iridate $Dy_2Ir_2O_7$ prepared by the CsCl flux method were investigated by means of specific heat and magnetization measurements. The low temperature behaviour was confirmed to be consistent with previously published results, whereas the high temperature data and their analysis are presented for the first time. At 128 K, a bifurcation of magnetization measured under zero field cooled and field cooled regimes coincides with an onset of the anomaly in specific heat pointing to an ordering of the iridium sublattice. Further, our high-temperature magnetization and specific heat data are interpreted in the frame of the previously determined crystal field scheme for $Dy_2Ir_2O_7$. The difference between calculations and experimental data, represented by 20 K specific heat anomaly, suggests the behaviour of studied iridate beyond the mean field approximation. This is in contrast with the rest of the heavy rare-earth pyrochlore iridates, where the mean field approximation constitutes an adequate description.

1. ## Introduction

The search for quantum spin liquid states (QSL) in geometrically frustrated magnets is traced back to seventies when Anderson (1) proposed the resonating valence bond state to describe magnetic moments on two-dimensional triangular lattice. A significant scientific attention has been attracted to spin liquids since eighties as QSL was suggested to be a key ingredient of the high-temperature superconductors' physics (2). Beside triangular lattice, the geometrical frustration possibly leading to QSL has been studied in other 2D and layered systems, e.g. kagomé lattice or honeycomb lattice (3). Analogously, QSL has been sought in three-dimensional frustrated lattices, represented mainly by the so called hyperkagomé and pyrochlore structures (4),(5). Focusing on the later structure (space group *F d -3 m*, 227), the structure in which here reported $Dy_2Ir_2O_7$ iridate crystallizes, it is formed by a network of interpenetrating corner-sharing tetrahedra with vertices occupied by rare-earth element (*A*) and d-element (*B*), respectively. $A_2B_2O_7$ pyrochlore structure with both $A^{3+}$ or/and $B^{4+}$ ions magnetic thus provides an ideal frame for emergence of quantum disordered ground states, or other exotic states (6),(7),(8).

$Dy_2Ir_2O_7$ is the iridium analogue of thoroughly studied $Dy_2Ti_2O_7$, a classical spin ice material with water-ice-like arrangement of two spins pointing in and two spins pointing out of their respective tetrahedra (2I2O) (9),(10). The spin ice ground state is highly degenerated and reveal corresponding non-zero entropy down to 0 K. However, $Dy_2Ti_2O_7$ was found to be beyond a simple nearest-neighbour dipole interaction, the simplest prerequisite for emergence of the spin-ice state, and rather revealed both ferromagnetic and antiferomagnetic dipole, and exchange interactions (11). It was suggested that $Ir^{4+}$ magnetic moments in $Dy_2Ir_2O_7$ are ordered antiferromagnetically with so called all-in-all-out (AIAO) ordering below temperature of magnetic ordering and/or metal-insulator transition, $T_{Ir}$ (= 134 K (12)), similarly to most of other rare earth iridate pyrochlores $A_2Ir_2O_7$ (13),(14),(15). The low-temperature anomaly (below 5 K), observed in all dc- and ac-susceptibility, and specific heat data (16), was ascribed to the Dy magnetic moments dynamics; no long-range ordering in $Dy_2Ir_2O_7$ was observed down to 0.1 K. However, very recent a study on $Dy_2Ir_2O_7$ (12) showed also a presence of long-range ordering between Dy magnetic moments. Several magnetic peaks described by AIAO structure were observed in neutron diffraction patterns. Simultaneously, a clear frequency development of low-temperature anomaly in ac-susceptibility data was followed, further supported by specific heat analysis. Hence, so called fragmented monopole crystal state was suggested to be realized in $Dy_2Ir_2O_7$ (12), that is, both antiferromagnetic order and a Coulomb phase spin liquid co-inhabit the system, as was previously reported for $Ho_2Ir_2O_7$ analogue (17). The competition between *A-A* interactions favouring a spin ice state, and the influence of the staggered field of iridium ions inducing AIAO ordering leads to the total free $Dy^{3+}$ ion moment being split into two halves (12). Nevertheless, considering previous results, especially the data measured on single crystalline sample (12), the role of Ir sublattice on Dy sublattice properties still requires further examination.

The present study reports on a complementary magnetization and specific heat measurements on a well-characterized $Dy_2Ir_2O_7$ polycrystalline sample (18). Presented high-temperature data, including, so-far unreported, specific heat data, bring additional information on paramagnetic regime of dysprosium sublattice subjected to staggered field induced by the iridium sublattice. The low-temperature data, although measured and analysed, are presented only briefly in order to minimize the overlap with recently published studies. The measured properties are discussed in the frame of crystal field (CF) scheme reported in Ref. (12), showing only a partial agreement between data and mean-field calculations.

2. **Sample preparation and characterization**

Bulk properties of $Dy_2Ir_2O_7$ were investigated on the polycrystalline sample synthesized by CsCl-flux assisted solid state reaction; the preparation route is described elsewhere (18). The structure parameters of the pyrochlore *F d -3 m* structure were determined employing powder X-ray diffraction. The lattice parameter and single free fraction coordinate of oxygen's Wyckoff position were refined to be $a$ = 10.192(2) Å and $x_{48f}$ = 0.334, respectively; well in agreement with previously determined values (12),(15),(19).

Further check of phase purity and homogeneity was performed using scanning electron microscope with back-scattered electron detector and energy dispersive X-ray analyzer (EDX). A typical bi-pyramid shape of prepared single crystals (of dimension ~ 10 μm$^3$) suggest a good quality material. Although EDX technique cannot determine an oxygen content in the sample, the Dy:Ir stoichiometry was confirmed to be 50(2):50(2).

The powder sample was cold-pressed into pellets, the pieces of ~ 10 mg and ~1 mg were used for measurements at high- and low-temperatures, respectively. The specific heat measurement at temperatures 0.4 – 300 K was performed using PPMS, Quantum design, equipped with the $^3$He insert. The magnetic susceptibility in the temperature range 1.8 – 330 K was measured employing superconducting quantum interference device magnetometer (MPMS7, Quantum design); the low-temperature data (down to 0.4 K) were collected using a PPMS equipped with $^3$He insert and custom-made Hall probes option (13),(20),(21).

### 3. Magnetization and specific heat

Temperature evolutions of magnetic susceptibility and specific heat in $Dy_2Ir_2O_7$ presented in Figure 1 are dominated, besides the low-temperature anomaly (presented in the inset of Figure 1b) previously thoroughly investigated and reported on elsewhere (12),(22), by a bifurcation of zero-field-cooled (ZFC) and field-cooled (FC) magnetization and a broad anomaly, respectively, at 128 K. In the context of other $A_2Ir_2O_7$ iridates (13),(15),(21),(23) the observed behaviour is ascribed to the magnetic ordering, either long-range, short-range, or both, of Ir sublattice, and metal to insulator transition. The temperature of a bifurcation, $T_{Ir}$ (= 128 K), is well in agreement with previous results (12),(15), and fits perfectly into the $A_2Ir_2O_7$ phase diagram presented in Ref. (13). $T_{Ir}$ marks the onset of the, so far unreported, anomaly in specific heat (vertical line in Figure 1), well in agreement with other pyrochlore iridates (13).

To properly analyse the anomaly we followed the approach used investigating other $A_2Ir_2O_7$ analogues (13), that is, the estimation of related entropy. The measured data were fitted by a third-order polynomial (blue line in Figure 1b) at highest temperatures to roughly estimate the specific heat without magnetic contribution of Ir sublattice. The two data sets were subsequently subtracted and their difference integrated to estimate the entropy of the anomaly $S_{Ir}$. The calculated value ($S_{Ir}$ = 1.26(3) JK$^{-1}$mol$^{-1}$) is in line with values reported for lighter $A_2Ir_2O_7$, $Sm_2Ir_2O_7$ and $Eu_2Ir_2O_7$, where Raman spectroscopy results pointed to a structural change at $T_{Ir}$ (15). Although the $S_{Ir}$ value is higher than for other heavy $A_2Ir_2O_7$ (13) and $Nd_2Ir_2O_7$ (15), it is still significantly smaller than the magnetic entropy expected for $Ir^{4+}$ with $S$ = 1/2. Similar values of estimated electronic specific heat coefficient $\gamma_{Ir} = S_{Ir}/T_{Ir}$ are followed within the rare-earth $A_2Ir_2O_7$ series (13), indicating non-metallic state; while $\gamma_{Ir}$ = 9.2(2) mJmol$^{-1}$K$^{-2}$ for $Dy_2Ir_2O_7$ points rather to the semi-metallic ground state, similarlyas in $Lu_2Ir_2O_7$ (23).

A bifurcation of ZFC and FC magnetization is traced in magnetic field up to 0.1 T (Figure 1a). Most of ZFC and FC magnetization curves monotonically increase with decreasing temperature, while ZFC magnetization in 0.005 T forms a bump and then increases with decreasing temperature. This observation can be attributed to the interactions between iridium moments and demonstrates itself also in the case of $Lu_2Ir_2O_7$ (23). No clear anomaly below $T_{Ir}$ is observed contrary to e.g. $Er_2Ir_2O_7$ and $Yb_2Ir_2O_7$ (13).

The magnetic susceptibility measured in high magnetic field up to 7 T (Figure 2) develops negligibly with applied field. Fitting the high temperature data, from 150 K to 330 K, to the Curie-Weiss law resulted in a paramagnetic Curie temperature $\theta_P$ = -13.1(2) K and an effective magnetic moment $\mu_{eff}$ = 10.2(1)$\mu_B$. Extrapolating the fit to lower temperatures, it overlaps with the measured data down to ~50 K (Figure 2). The value of $\theta_P$ fits perfectly into the trend indicated by other heavy rare-earth $A_2Ir_2O_7$ (13), that is, $\theta_P$ decreases with increasing ionic radius of $A^{3+}$. The small negative value of $\theta_P$ points to rather weak antiferromagnetic correlations among $Dy^{3+}$ ions, which is in agreement with proposed fragmented state scenario (12). Fitted effective moment is close to $\mu_{eff}$ of $Dy^{3+}$ free ion (= 10.65 $\mu_B$). Roughly correcting the data for the magnetic contribution of Ir sublattice to total signal, we subtracted the $Lu_2Ir_2O_7$ analogue data (23) from $Dy_2Ir_2O_7$, see Figure 2, leading to somewhat lower values of fitted parameters: $\theta_P$ = -11.3(2) K and $\mu_{eff}$ = 8.9(1) $\mu_B$.

The isothermal magnetization measured at low temperatures reveals magnetic moment of the system reaching the value of 4.8 $\mu_B$ in 7 T (Figure 3). Further increase of magnetization in higher fields is expected, however a saturated value can be estimated to 5 $\mu_B$ or only slightly higher value, that is, half of the value expected for the $Dy^{3+}$ free ion ($\mu_{Dy3+}$ = 10 $\mu_B$). Such an observation is in a good agreement with neutron diffraction results and fragmented state scenario reported recently by Cathelin et al. (12).

4. Discussion – crystal field calculations

Utilizing the crystal field parameters determined from inelastic neutron scattering data (12), we calculated, employing in-house computer codes (recently utilized and described in Refs. (24),(21)), isothermal magnetization curves presented in Figure 3. The calculations give significantly lower magnetization compared to experimental one, which is, nevertheless, accounted for by a contribution of the Ir sublattice to total magnetization. The magnetic moment of the $Ir^{4+}$ free ion with S = 1/2 reaches $\mu_{Ir}$ = 1.74 $\mu_B$. Analogous experimental data, reaching about $\mu = <\mu_{A3+}>_{111}+ \mu_{Ir}$ value in high magnetic field, were collected also for other $A_2Ir_2O_7$ (13), demonstrating a very similar impact of the Ir sublattice magnetism on total magnetic properties of the compounds.

To further confront the microscopic data (12) to measured bulk properties, the mean-field calculations of magnetic susceptibility from reported CF parameters were done. The calculations indicate a strong anisotropy in the system; the susceptibilities along (∥) and perpendicular to (⊥) the local crystallographic <111> axes of the pyrochlore tetrahedron are presented in Figure 2. A relatively good agreement of experimental data and calculated curves, including the powder averaged calculated susceptibility, is observed only at highest temperature. The low-temperature part is reproduced by the CF calculations very poorly, indicating the behaviour beyond the mean-field approximation. The key feature of this behaviour is the Ising character of the ground state doublet well isolated (around 29 meV from first excited doublet) as was found in Ref. (12). We highlight that the agreement between data and calculations was significantly better for previously studied heavy rare-earth $A_2Ir_2O_7$ (13),(21).

Finally, the CF energy scheme (12) was utilized to calculate Schottky contribution to specific heat, $C_{Schottky}$, see Figure 1b. To obtain an estimate on the magnetic specific heat related to the Dy sublattice, the $Lu_2Ir_2O_7$ specific heat, $C_{Lu}$, was subtracted from the measured data, $C_p$, simulating all magnetic contribution of the Ir sublattice, $C_{Ir}$, and electron, $C_{el}$, and phonon, $C_{ph}$, contributions. Both, $Dy_2Ir_2O_7$ and $Lu_2Ir_2O_7$, data sets were first fitted by the polynomial at high temperatures to eliminate the contribution of specific heat anomalies at slightly different temperatures (Figure 1b), and subsequently subtracted. The obtained magnetic contribution, $C_{Dy}$, is characterized by three anomalies. The lowest-temperature anomaly (inset of Figure 1b) was discussed in detail in Ref. (12) and is supposed to origin in the magnetic interactions between $Dy^{3+}$ ions. The broad high-temperature anomaly is well described by $C_{Schottky}$ (Figure 1b), and is thus unambiguously ascribed to the CF. The last anomaly, centred around 25 K, remains enigmatic. Of course, first explanation of the anomaly could be found in the usage of Lu$_2$Ir$_2$O$_7$ data as 'non-magnetic' analogue. Supposing (i) the $C_{Ir}$ is almost identical within the series, which is, to a certain degree, confirmed investigating the specific heat of other $A_2$Ir$_2$O$_7$ (13),(21); and (ii) the $C_{el}$ is similar for all the rare-earth analogues; the $C_{ph}$ contribution, dependent specially on atomic properties of Dy and Lu, could be responsible for observed anomaly. Previously used Eu$_2$Ir$_2$O$_7$ data represent a better choice to obtain $C_{Dy}$ (12),(16). Unfortunately, previous reports present the $C_{Dy}$ only up to 10 K. Nevertheless, comparing the Eu$_2$Ir$_2$O$_7$ (15) and Lu$_2$Ir$_2$O$_7$ data (23), no significant difference, focusing mainly on relevant temperature range, is observed. Considering the anomaly intrinsic the explanation could be found in the low-energy high phonon density of states observed in neutron scattering data (12). However, further data on energy spectrum of Lu$_2$Ir$_2$O$_7$ or Eu$_2$Ir$_2$O$_7$ are essential to confirm/disprove such explanation.

All in all, the high-temperature magnetization and specific heat characteristics, on one hand, well fit into the phase diagram and evolution of physical properties of heavy rare-earth $A_2$Ir$_2$O$_7$ with $A$, that is, with a change of interatomic distances and angles substituting $A$ element within the pyrochlore lattice. On the other hand, the agreement of data and previously reported CF model (12) is relatively weak, indicating other mechanisms, including exchange interactions, frustration, and possibly also Ir sublattice exchange field, play a significant role in a formation of magnetic, and generally electronic, properties of Dy$_2$Ir$_2$O$_7$. The suggested fragmented state in the iridate (12) could explain the observed behaviour, nonetheless, further, especially microscopic, studies are highly desirable.

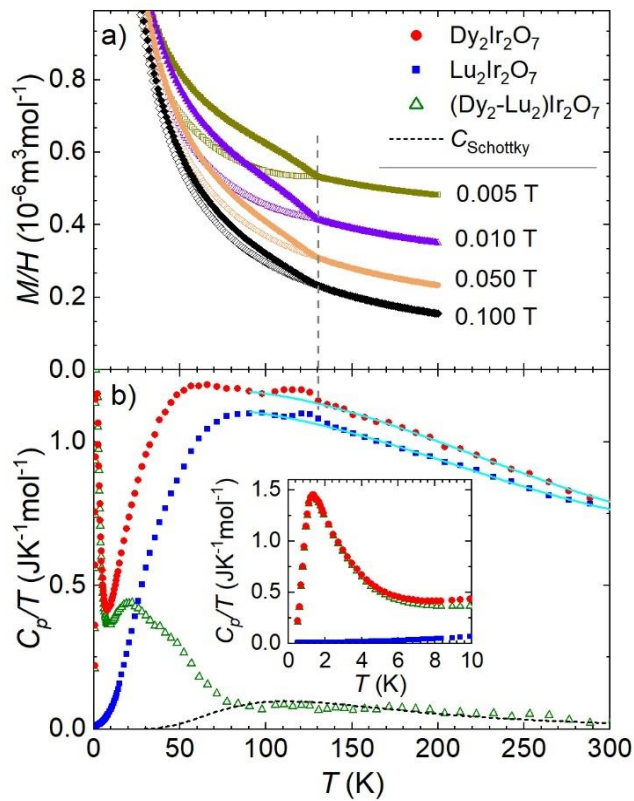

Figure 1. a) magnetic susceptibility measured under zero field cooled (open symbols) and field cooled (full symbols) conditions in various external magnetic fields. b) specific heat of $Dy_2Ir_2O_7$, $Lu_2Ir_2O_7$ (23), and their difference. Schottky contribution to specific heat calculated from CF energies is plotted as well. Blue lines are the polynomial fits of the high-temperature specific heat above and just below the iridium sublattice-connected anomaly. Vertical line marks the temperature of a bifurcation of ZFC and FC data. The inset contains zoomed out low-temperature anomaly in specific heat.

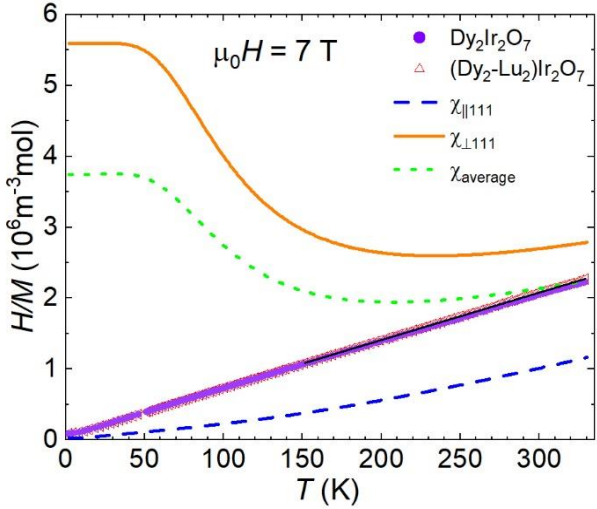

Figure 2. Inverse magnetic susceptibility measured in external magnetic field of 7 T. The $Dy_2Ir_2O_7$ data, as well as difference $(Dy_2-Lu_2)Ir_2O_7$ data are shown; the data overlap significantly. The fit to Curie-Weiss behaviour (black line) in the temperature range 150 – 330 K is plotted. Orange and blue lines represent calculated susceptibilities along and perpendicular to local <111> crystallographic directions, respectively, while green dotted line stands for their powder average.

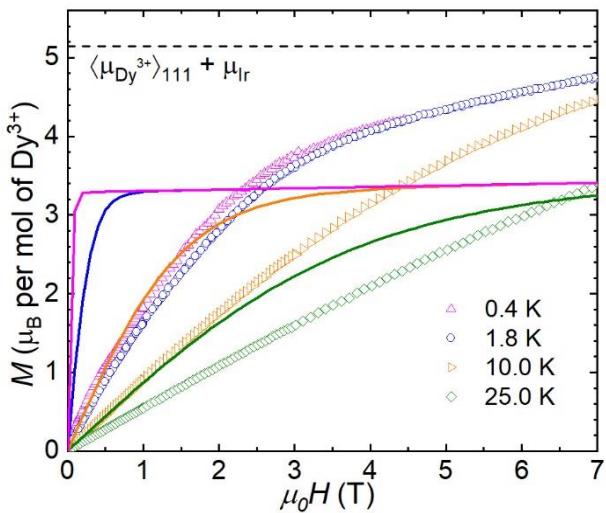

Figure 3. Isothermal magnetization of $Dy_2Ir_2O_7$ at selected temperatures. The powder averaged values of isothermal magnetization calculated from CF parameters (of $Dy^{3+}$) (12) are shown as full lines of corresponding colors. The expected saturation value of magnetization $\mu = \langle\mu_{A^{3+}}\rangle_{111} + \mu_{Ir}$ is represented by horizontal dashed line.

## 5. Summary

Dy$_2$Ir$_2$O$_7$ was prepared by solid-state reaction mediated by the CsCl-flux, and studied by means of magnetization and specific heat measurements. The ordering of the Ir sublattice was followed on both data sets below $T_{Ir}$ = 128 K. Negative paramagnetic Curie temperature, $\theta_P$ = -11.3(2) K, suggested only relatively weak antiferromagnetic correlations between Dy$^{3+}$ moments, following, nevertheless, the evolution of $\theta_P$ with $A$ in heavy rare-earth $A_2$Ir$_2$O$_7$. The measured magnetic moment reached about half of the value of the Dy$^{3+}$ free ion under the field of 7 T, well in agreement with previously published data and proposed fragmentation model. The exotic ground state/behaviour of this iridate was further supported by a poor agreement between measured data and from CF parameters calculated magnetization and Schottky specific heat.

**Acknowledgement**

The preparation, characterization and measurement of bulk physical properties on Dy$_2$Ir$_2$O$_7$ sample were performed in MGML (http://mgml.eu/), which was supported within the program of Czech Research Infrastructures (project no. LM2018096). Additionally, this work was supported by the Czech Science Foundation under Grant No. 18-09375Y. The work of K.V. was further supported by GAUK Project Number 558218.